\documentclass[reprint,twocolumn,superscriptaddress,showpacs,nofootinbib,notitlepage]{revtex4-1}

\usepackage{graphicx}
\usepackage{latexsym,amsmath,amssymb,lmodern,float,url}
\usepackage{natbib}
\usepackage{color}
\usepackage{microtype}
\usepackage{slashed}
\usepackage{multirow}
\usepackage{subcaption}

\newcommand{\be}{\begin{equation}}
\newcommand{\ee}{\end{equation}}
\newcommand{\bea}{\begin{eqnarray}}
\newcommand{\eea}{\end{eqnarray}}
\def\beqs#1\eeqs{\be\begin{split} #1 \end{split}\ee}
\newcommand{\eq}[1]{Eq.~(\ref{eq:#1})}

\newcommand{\nn}{\nonumber}
\newcommand{\ba}{\begin{array}}
\newcommand{\ea}{\end{array}}

\usepackage[colorlinks=true,backref=false, linktocpage=true,
citecolor=blue,urlcolor=blue,linkcolor=blue,pdfpagemode=UseOutlines]{hyperref}

\hypersetup{%
  bookmarksnumbered=true,
  pdftitle = {},
  pdfsubject = {},
  pdfauthor = {},
  pdfkeywords = {}
}

\def\Re {\operatorname{{Re}}}
\def\Im {\operatorname{{Im}}}
\def\Tr {\operatorname{{Tr}}}

\newcommand{\asi}{\langle \sigma\rangle}
\newcommand{\chicon}{\langle \bar{\psi}\psi\rangle}

\renewcommand{\d}{\mathrm d}

\begin{document}
\title{Fermions at Finite Density in (2+1)d with Sign-Optimized Manifolds}

\author{Andrei Alexandru}
\email{aalexan@gwu.edu}
\affiliation{Department of Physics, The George Washington University, Washington, D.C. 20052, USA}
\affiliation{Department of Physics, University of Maryland, College Park, MD 20742, USA}
\author{Paulo F. Bedaque}
\email{bedaque@umd.edu}
\affiliation{Department of Physics, University of Maryland, College Park, MD 20742, USA}
\author{Henry Lamm}
\email{hlamm@umd.edu}
\affiliation{Department of Physics, University of Maryland, College Park, MD 20742, USA}
\author{Scott Lawrence}
\email{srl@umd.edu}
\affiliation{Department of Physics, University of Maryland, College Park, MD 20742, USA}
\author{Neill C. Warrington}
\email{ncwarrin@umd.edu}
\affiliation{Department of Physics, University of Maryland, College Park, MD 20742, USA}
\date{\today}

\newcommand{\Seff}{S_{\text{eff}}}
\begin{abstract}
We present Monte Carlo calculations of the thermodynamics of the (2+1) dimensional Thirring model at finite density. We bypass the sign problem by deforming the domain of integration of the path integral into complex space in such a way as to maximize the average sign within a parameterized family of manifolds. We present results for lattice sizes up to $10^3$ and we find that at high densities and/or temperatures the chiral condensate is abruptly reduced.
\end{abstract}

\maketitle


Monte Carlo methods are critical to the nonperturbative study of strongly interacting quantum field theories and many-body systems. In the lattice field theory approach, one discretizes spacetime and formulates observables as high dimensional lattice path integrals. For systems in thermal equilibrium, such integrals take the form $\langle \mathcal{O}\rangle = Z^{-1}\int{\mathcal DA~\text{e}^{-S}\mathcal{O}}$ where $Z$ is the partition function and $S$ is the (Euclidean) action. Path integrals are typically only computable by importance sampling, which relies on interpreting $\text{e}^{-S}/Z$ as a probability distribution. However, many theories of interest have complex actions. This {\em sign problem} is a major roadblock to the ab-initio study of such systems, including fermions at finite density.

For systems with complex actions $S=S_R + i S_I$, a common method is to sample according to the distribution $\text{Pr}(A) \sim \text{e}^{-S_R(A)}$, and to express observables as $\langle \mathcal{O}\rangle = \langle \mathcal{O}~\text{e}^{-i S_I}\rangle_R/{\langle \text{e}^{-i S_I} \rangle_R } $, where $\langle \cdot \rangle_R$ means averaging with respect to the $S_R$. This ``reweighting" procedure is effective if the average sign $\asi \equiv {\langle \text{e}^{-i S_I} \rangle_R } $ is not too small. However, $\langle \sigma\rangle$ typically decreases exponentially with the spatial volume $L^d$, chemical potential $\mu$ and inverse temperature $\beta$, so for cold dense matter standard reweighting fails~\cite{Gibbs:1986ut}. In response to this failure, many ideas have been explored: the complex Langevin~\cite{Aarts:2008rr}, the density of states method  \cite{Langfeld:2016mct}, canonical methods~\cite{Alexandru:2005ix,deForcrand:2006ec}, reweighting methods~\cite{Fodor:2001au}, series expansions in the $\mu$~\cite{Allton:2002zi}, fermion bags~\cite{Chandrasekharan:2013rpa}, and analytic continuation from imaginary $\mu$~\cite{deForcrand:2006pv}.

In a recently developed family of approaches to taming the sign problem, the original domain of integration $\mathcal{M}_O$ of the path integral is deformed to a submanifold $\mathcal{M}$ of the complexified field space.  A multidimensional generalization of Cauchy's integral theorem guarantees, for suitable deformations, that integrals of holomorphic functions (e.g. physical observables) remain unchanged. In contrast, integrals of non-holomorphic functions such as $\asi$ depends on $\mathcal{M}$, and therefore a judicious choice of manifold can increase $\asi$ and render reweighting feasible.

The first manifolds suggested were sets of multidimensional stationary phase contours called {\em Lefschetz thimbles}, $\mathcal{M}_L$ ~\cite{Cristoforetti:2012su,Cristoforetti:2013wha,Cristoforetti:2013qaa,Scorzato:2015qts}.  Analytically, $\mathcal{M}_L$ have been found in only a handful of cases which include  few dimensional integrals and quantum mechanical models~\cite{Tanizaki:2014tua,Kanazawa:2014qma,Fujii:2015bua,Tanizaki:2015rda}. Numerous algorithms have been developed to integrate on $\mathcal{M}_L$, but these methods have difficulty addressing which set of thimbles reproduce the results on $\mathcal{M}_O$~\cite{Mukherjee:2013aga,Fujii:2013sra,Alexandru:2015xva,DiRenzo:2015foa,Fukushima:2015qza,DiRenzo:2017igr,Bluecher:2018sgj}.  To address this, a {\em generalized thimble method} was developed. In this approach, one deforms $\mathcal{M}_O$ via the {\em holomorphic gradient flow} for a fixed time $T$, which yields a manifold $\mathcal{M}_T$ that approaches $\mathcal{M}_L$ as $T\rightarrow\infty$~\cite{Alexandru:2015sua}. The generalized thimble method has been applied to analyze bosonic and fermionic systems at finite density ~\cite{Alexandru:2016san,Nishimura:2017vav,Alexandru:2016ejd,Fukuma:2017fjq,Alexandru:2017oyw}, real-time linear response ~\cite{Alexandru:2016gsd,Alexandru:2017lqr}, and gauge theories~\cite{Alexandru:2018ngw}. One drawback to the generalized thimble method is it requires a computationally expensive Jacobian related to the manifold parametrization. This lead to developments in rapidly computable estimators~\cite{Alexandru:2016lsn} and in applying machine learning to approximate the manifold~\cite{Alexandru:2017czx}.  

To avoid all these difficulties, the sign-optimized manifold method was introduced in~\cite{Alexandru:2018fqp} wherein one deforms $\mathcal{M}_O$ to a manifold $\mathcal{M}_S$ that maximizes $\asi$ within a family of manifolds $\mathcal{M}_\lambda$ parameterized by a set of real numbers ${\lambda_i}$.  A similar method is described in~\cite{Mori:2017pne}. To guarantee that the path integral remains invariant under the deformation to $\mathcal M_S$, it is sufficient that $\mathcal{M}_O$ is continuously deformable to $\mathcal M_\lambda$ without crossing any singularity of the integrand. These conditions are satisfied by construct in our family of manifolds because our deformations are smooth and involve only finite shifts of the fields in the imaginary direction.

In this letter, we explore the finite density phase diagram of the two flavor $(2+1)$d Thirring model using the sign-optimized manifold method, extending the range in $(T,\mu)$ space beyond what is possible on $\mathcal{M}_O$.  

We parameterize the manifold $\mathcal M_\lambda$ by its projection on the real space $\mathcal{M}_O$, so that integration on $\mathcal M_\lambda$ may be achieved by integrating on $\mathcal{M}_O$ with the inclusion of a Jacobian, which is included into an effective action. Thus, the expectation value of an observable $\mathcal{O}$ can be written as:
\begin{equation}\label{eq:expectation-parameterized}
\left<\mathcal O\right>=
\frac
{\int_{\mathcal{M}_O}\mathcal DA\;\mathcal O[\tilde A(A)] e^{-\Seff[A;\lambda]}}
{\int_{\mathcal{M}_O}\mathcal DA\; e^{-\Seff[A;\lambda]}},
\end{equation}
where $\tilde A(A)$ is the point on the manifold $\mathcal M_\lambda$ parametrized by $A$, $\Seff\equiv S-\ln\det J$ is the effective action, and $J$ is the Jacobian of the parametrization.  The average sign on $\mathcal M_\lambda$ is given by
\begin{equation}\label{eq:sign1}
\left<\sigma\right>_{\lambda}
=
\frac
{\int_{\mathcal{M}_O} \mathcal DA\;e^{-\Seff[A;\lambda]}}
{\int_{\mathcal{M}_O} \mathcal DA\;e^{-\Re\Seff[A;\lambda]}}.
\end{equation} 

The numerator of Eq.~(\ref{eq:sign1}) is independent of $\lambda$ because it is the integral of a holomorphic function in $A$, but the denominator depends on $\lambda$ because $e^{-\Re\Seff}$ is not holomorphic.  We are interested in maximizing this as a function of the manifold parameters $\lambda$ --- this is equivalent to maximizing $\log \left|\left<\sigma\right>_\lambda\right|$. The gradient of $\log \left|\left<\sigma\right>_\lambda\right|$ with respect to $\lambda$ is  
\begin{align}
&\nabla_\lambda \log \left|\left<\sigma\right>_{\lambda}\right|
=\nn\\&
\frac
{\int_{\mathcal{M}_O} \mathcal DA\;e^{-\Re \Seff[A;\lambda]}\left[\nabla_\lambda S_R - \Re \Tr J^{-1} \nabla_\lambda J\right]}
{\int_{\mathcal{M}_O} \mathcal DA\;e^{-\Re \Seff[A;\lambda]}}\,.
\end{align}                                                                                                                      
This gradient is the phase-quenched expectation value $\left<\nabla_\lambda \Seff \right>_{\Re \Seff}$, and is therefore free from a sign problem. This allows $\nabla_\lambda \log \left|\left<\sigma\right>_{\lambda}\right|$ to be computed reliably by a short Monte Carlo simulation at each gradient ascent step. To do gradient ascent we use the Adaptive Moment Estimate algorithm~\cite{2014arXiv1412.6980K}. We stress that the sign-free nature of the calculation of the gradient is central to the method and allows our calculations to be efficient even when $\asi$ is exponentially small.


One potential issue is that the computation of $\det J$ is an expensive operation --- for general $J$, this requires time proportional to the cube of the spacetime volume. In~\cite{Alexandru:2018fqp} it was shown that this computational cost can be avoided by proposing a family of manifolds for which the Jacobian matrix is diagonal. We use a similar family here (details below). A more general ansatz with non-diagonal Jacobian with nearest-neighbor correlations has been shown to improve the sign problem in bosonic theories, with increased computational expense~\cite{Bursa:2018ykf}.

To integrate on our curved manifolds, we have implemented a modified version of hybrid Monte Carlo (HMC).  We define a Hamiltonian
 \begin{equation}
 \label{eq:hamiltonian}
 H(\pi,A) = \frac{1}{2}\sum_x{\pi_x[J(A)J^{\dagger}(A)]^{-1}_{xy}\pi_y}+S_R(\tilde{A}(A))
 \end{equation}
 and sample according to the distribution $P(\pi,A) \sim \text{e}^{-H(\pi,A)}$. Marginalizing over the momenta yields the distribution $P(A) \sim |\text{det}J(A)|\text{e}^{-S_R(A)}$. Sampling according to $P(A)$ then reweighting with the residual phase $\text{e}^{-i\Im \Seff}$ yields the correct observables.  For generic dense Jacobians the derivatives $\partial H/\partial A_x$ are extremely expensive to compute, but for manifolds with diagonal Jacobians the derivatives are computed analytically and implemented with sparse matrices. Thus, HMC allows for sampling on $\mathcal{M}_\lambda$ as fast as sampling on $\mathcal{M}_O$. Due to this Jacobian's structure the evolution of Eq.~(\ref{eq:hamiltonian}) can be calculated with implicit and explicit symplectic integrators. Both were implemented and found to agree.
 
 We now apply the sign-optimized manifold method to the $(2+1)$d Thirring model defined by the lattice action
\begin{equation}\label{eq:lattice-action}
S=\sum_{x,\nu} \frac{N_F}{g^2} (1-\cos A_\nu(x))+\sum_{x,y} \bar\psi^a(x) D_{xy}(A)  \psi^a(y)
\end{equation}
where $-\pi<A_\mu(x)\le\pi$ is a compact bosonic auxiliary field ~\cite{DelDebbio:1997dv,Hands:1999id,Christofi:2007ye}.  By virtue of the compact fields, $\mathcal{M}_O=(S^1)^N$ and the deformed manifold are submanifolds in the complexified space $(S^1\times \mathbb{R})^N$.  The staggered fermion matrix is given by
\begin{align}
D_{xy} = m\delta_{xy} + \frac{1}{2}\sum_{\nu=0}^2  
\Big[& 
 \eta_{\nu}(x) e^{i A_\nu(x)+\mu \delta_{\nu 0}} \delta_{x+\hat\nu, y}\nn\\
 &-\eta^\dag_{\nu}(y)e^{-i A_\nu(y)-\mu \delta_{\nu 0}}  \delta_{x, y+\hat\nu}
\Big], \nn
\end{align} 
where $\eta_\nu(x)=(-1)^{x_0+\ldots+x_{\nu-1}}$,
the flavor indices $a$ taking values from $1,\ldots,N_F/2$, $g$ is the coupling, and $m$ is the bare mass.  
There are different lattice actions which naively appear to have as their continuum limit the $(2+1)$-dimensional Thirring model. 
A substantial literature exists studying different discretizations of the $(2+1)$-dimensional Thirring model at zero density, with emphasis on determining the critical $N_F$ below which the chiral condenstate $\chicon$ is nonzero when $m\rightarrow0$~\cite{DelDebbio:1997dv,Hands:1999id,Christofi:2007ye,Gies:2010st,Janssen:2012pq,Wellegehausen:2017goy}. It is however, unclear which discretizations are equivalent in the continuum limit. For our purpose the action in Eq.~(\ref{eq:lattice-action}) defines what we mean by Thirring model.

Integrating out the fermions in \eq{lattice-action} gives
\begin{equation}\label{eq:action}
S=N_F 
\left(  
\frac{1}{g^2}\sum_{x,\nu} (1-\cos A_\nu(x)) - \frac12\log\det D(A)
\right)
\text.
\end{equation}

\begin{figure*}[t]
 \includegraphics[width=0.49\textwidth]{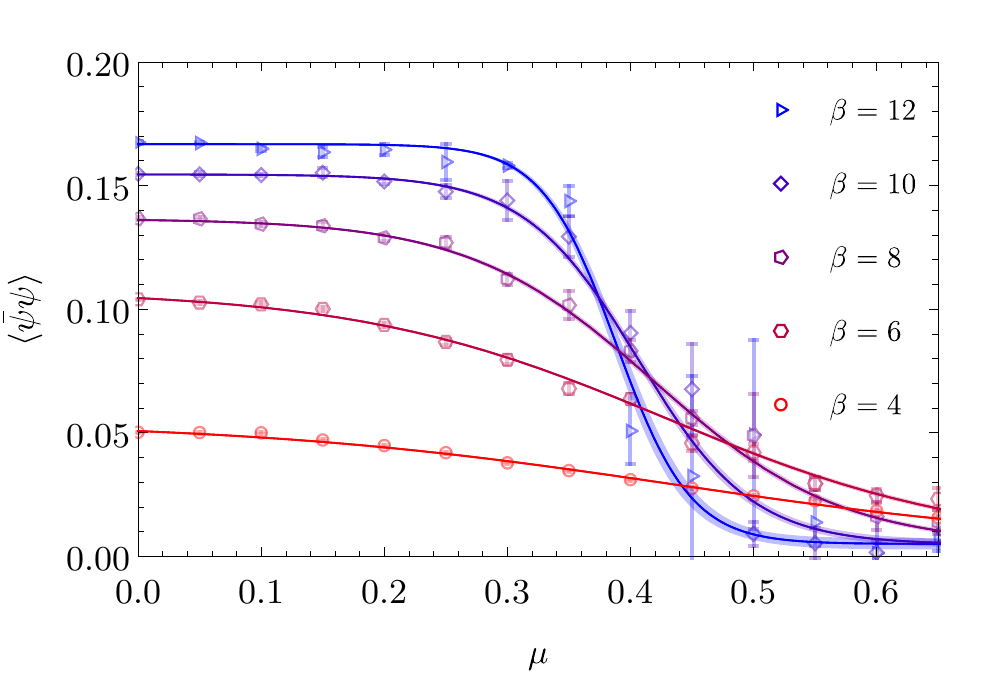}
\includegraphics[width=0.47\textwidth]{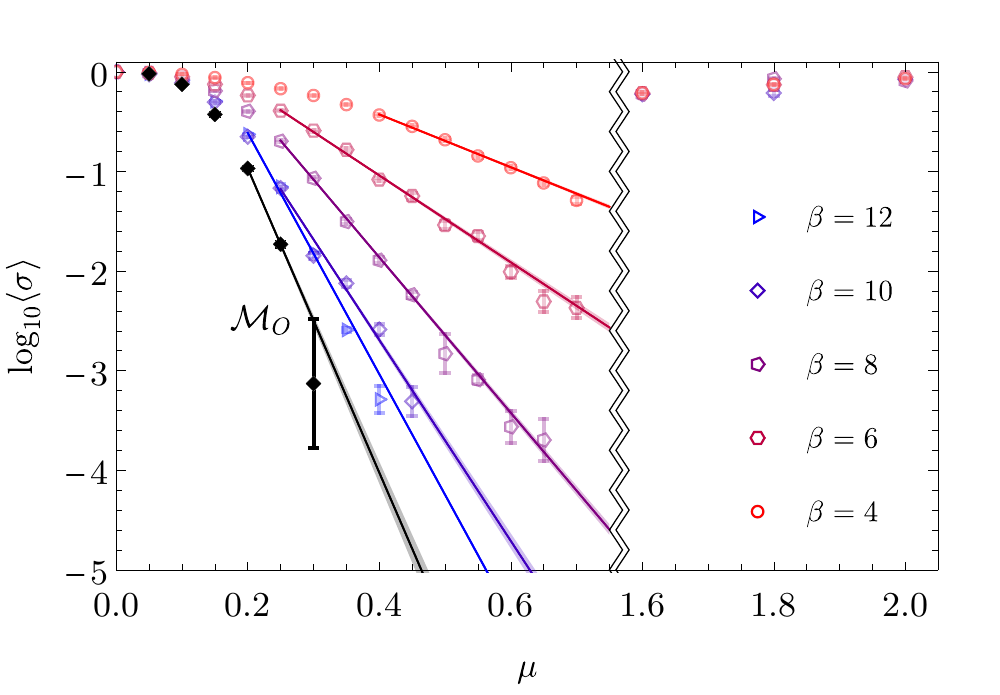}
\caption{ $\chicon$ (left) and $\langle\sigma\rangle$ (right)  as a function of $\mu$ for $\beta\times 6^2$ lattices. Notice the increase of $\langle\sigma\rangle$ for large values of $\mu$ as expected from the discussion in the text.  The black points are $\asi$ for simulations on $(S^1)^N$ on a $\beta=10$ lattice.\label{plt:6}}
\end{figure*}

We presently study the phase diagram in the $(T,\mu)$ plane for $N_F = 2$. For $\mu \ne 0$, the determinant $\det D(A)$ is complex and we must address the resulting sign problem.  

For insight into a family of manifolds which may increase $\asi$, we look to the $\mu\rightarrow\infty$ limit of the theory. In this limit, the density matrix is dominated by forward time links, and the path integral becomes
\begin{equation}
	\label{eq:factorize}
	Z = \left[\int \d^3 A \;e^{\frac 1 {g^2}\left(\sum_\nu \cos A_\nu\right) + \mu + \frac 1 2 i A_0}\right]^{\beta V}
\end{equation}
where only the leading terms in $e^{\beta\mu}$ are included. In this limit, the path integral factorizes, and the sign problem itself comes only from the integral over $A_0$.  Consequently, we will consider $\mathcal{M}_\lambda$ in which $A_1$ and $A_2$ remain on  $\mathcal{M}_O$, and $\Im \tilde A_0(x)$ depends only on $A_0(x)$,  not on any other link. Such factorizable manifolds have the desirable property that $J$ is diagonal.

At weak coupling ($g^2\rightarrow 0$) one expects the partition function to be dominated by the saddle point with the smallest action, which is $A_0(x) = i \alpha,~A_1(x)=A_2(x)=0$ for all $x$. As found in lower dimensional Thirring models the thimble attached to this critical point can be approximated by a shift of fields in the imaginary direction. This suggests that a shift $A_0(x) \rightarrow A_0(x) + i \alpha$ will improve $\asi$, and this was confirmed in simulations~\cite{Alexandru:2015xva,Alexandru:2015sua,Alexandru:2016ejd}. 

Consistent with these observations, we extend the manifolds used in~\cite{Alexandru:2018fqp} to the following three-parameter family:
\beqs
\tilde A_0 = A_0 + &i(\lambda_0 + \lambda_1 \cos A_0+ \lambda_2\cos(2 A_0))\text,\\
&\tilde A_1 = A_1\;\text,\;\tilde A_2 =A_2\text.
\label{eq:ansatz}
\eeqs
Every member of the family of manifolds above can be smoothly deformed to $(S^1)^N$ with the interpolation $\left(\tilde A_0\right)_t = A_0 + i t \left(\lambda_0 + \lambda_1 \cos A_0+ \lambda_2\cos(2 A_0)\right)$ with $0\leq t \leq1$ shows. Moreover, the imaginary shift is bounded, so the  condition for the applicability of Cauchy's theorem is satisfied. 
   
The results presented use bare parameters $g=1.08$ and $m=0.01$.
We quote the results of our simulations using lattice units. 
To demonstrate that we are in the strong coupling regime and to ascertain whether we are not too far from the continuum and thermodynamic limits, we measure the mass of the lowest fermionic and bosonic excitations 
by fitting the large-time behavior of correlators $\left<\mathcal O_f(t) \mathcal O_f(0)^\dagger\right>$ and $\left<\mathcal O_b(t) \mathcal O_b(0)^\dagger\right>$, where $\mathcal O_f(t) = \sum_{\vec x}\psi(t, \vec x)$ and $\mathcal O_b(t) = \sum_{\vec x}(-1)^{x_0+x_1+x_2}\bar\psi\psi(t, \vec x)$. Using a spatial volume of $L^2=10^2$ we find $m_f = 0.46(1)$ and $m_b = 0.21(1)$. The masses depend slightly on $L^2$,
but in all cases $m_b / m_f \ll 2$. This indicates that the system is strongly coupled since 
the binding energy of the boson is comparable to the $2m_f$. 

In this work, we calculate on six lattice geometries. We perform a series of simulations with fixed volume $L^2=6^2$ and varying temperature $\beta= 4,6,8,10,12$ to scan the $(T,\mu)$ plane, and we perform one simulation with $L^2=10^2$ and $\beta=10$ to investigate
the finite volume effects. The parameters $\lambda_i$ are typically smooth functions of $\mu$.  For the $12\times 6^2$ lattice with $\mu = 0.30$ as an example, we found $\lambda_0= 0.218, \lambda_1 = -0.126, \lambda_2=0.042$. 
On $\mathcal M_S$'s, we performed Monte-Carlo calculations generating between $10^2$ and $10^8$ independent configurations depending on the magnitude of $\asi$. 

\begin{figure*}[t!]
\includegraphics[width=0.49\textwidth]{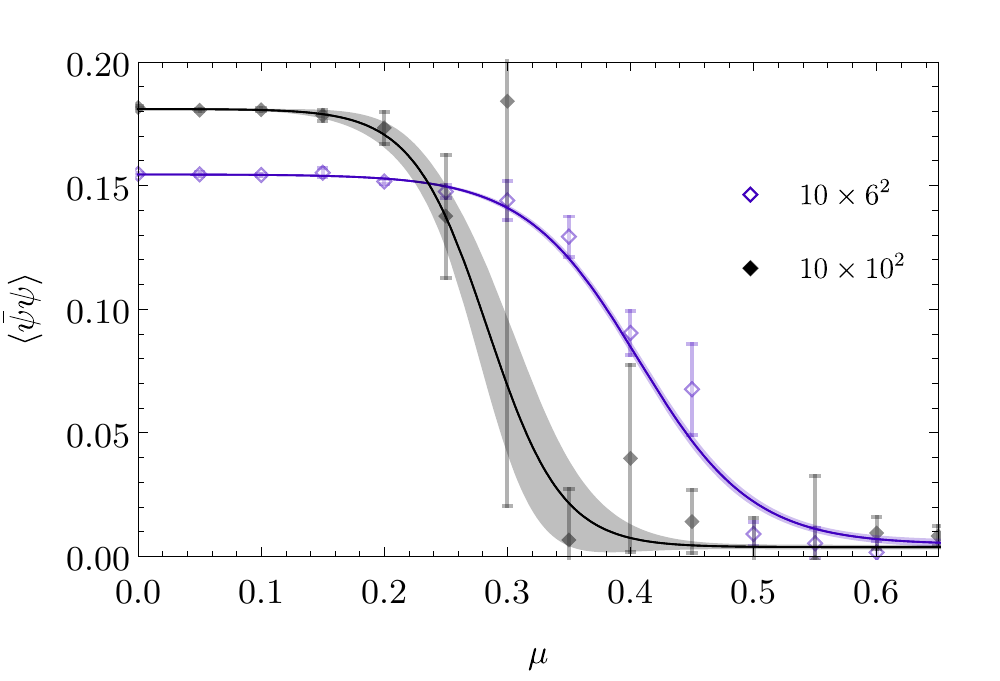}
\includegraphics[width=0.49\textwidth]{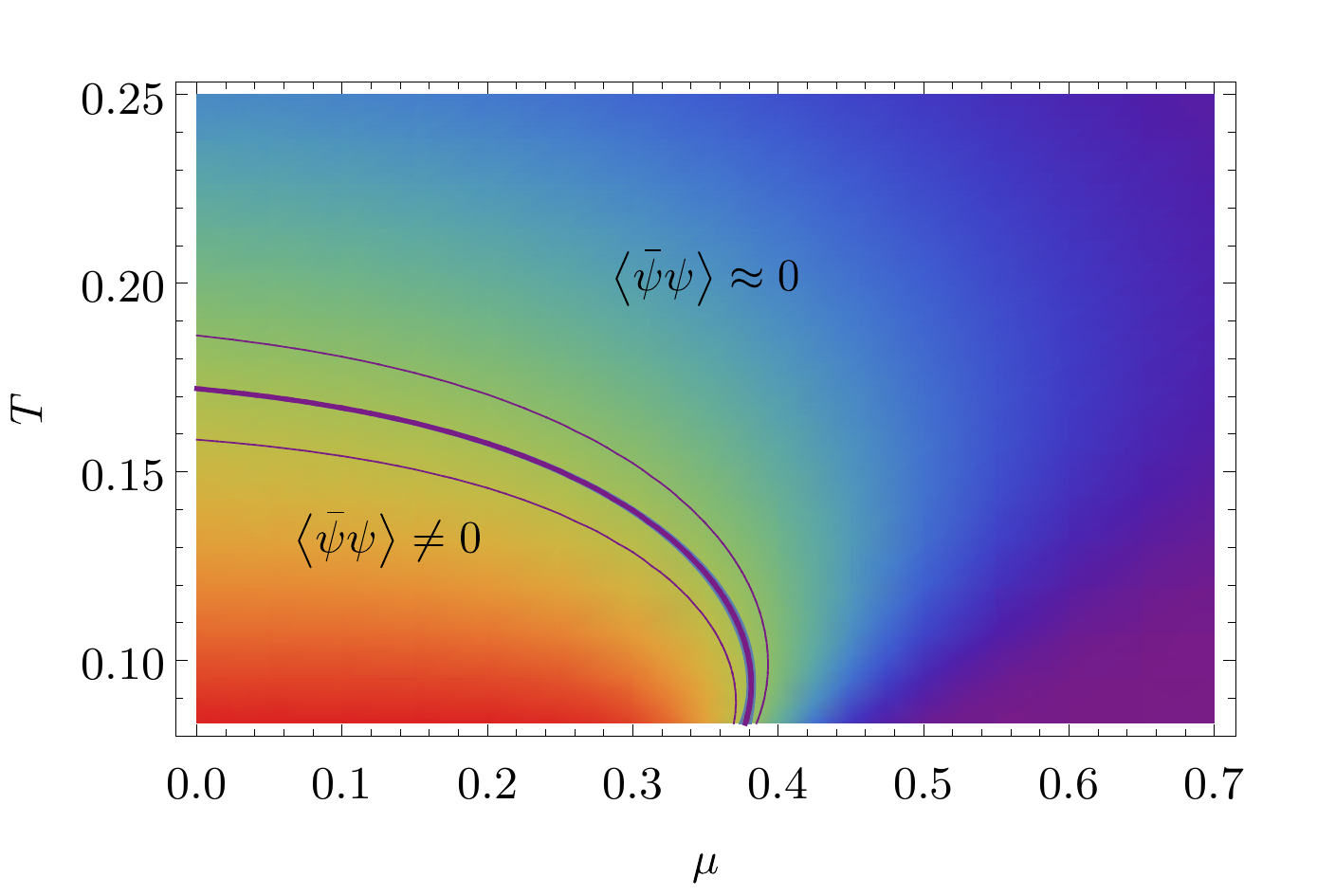}
\caption{ Left: $\chicon$ as a function of $\mu$ for $\beta=10$ at two different volumes: $L^2=10^2$ and $6^2$. A sharpening of the chiral transition can be seen as the volume is increased.
Right: $\chicon$ as a function of $T$ and $\mu$ for a spatial volume of size $6^2$. The thick central band indicates the location of $\chicon_{\mu,T}=0.5\chicon_{0}$ and its width represents the statistical error. The peripheral thin lines indicate $\chicon_{\mu,T}=(0.5\pm 0.05)\chicon_{0}$ to help gauge the sharpness of the transition.
\label{plt:7}}
\end{figure*}

The advantages of using $\mathcal{M}_S$ over a naive calculation on $(S^1)^N$ can be ascertained by computing $\asi$. When computed on $(S^1)^N$, $\asi$ decreases (exponentially) with $\mu$. On $\mathcal{M}_S$, $\asi$ initially decreases, but near saturation it increases and approaches unity, as can be seen on Fig.~\ref{plt:6}. This is consistent with expectations due to the discussion of limiting behavior around Eq.~(\ref{eq:factorize}).

In order to quantify the speedup gained on $\mathcal{M}_S$, note that
 the number of measurements required for a fixed precision scales like $\asi^{-2}$. Thus the speedup may be estimated by computing $\asi_{\mathcal{M}_S}^2/\asi_{(S^1)^N}^2$. Computing this ratio is difficult however because $\asi_{(S^1)^N}$ is very small at large $\mu$. We therefore estimate the value of $\asi_{(S^1)^N}$ by performing a fit to $\log \asi$ (see Fig.~\ref{plt:6}). Using this fit, we can compare the $\asi$ at large $\mu$. We find that on a $\beta\times L^2=10\times 6^2$ lattice for $\mu=0.45$, $\asi_{\mathcal{M}_S}^2/\asi_{\mathbb R^N}^2\approx10^4$, indicating a sizeable speedup.
 
 All $\chicon(\mu)$ are fit well with the ansatz: $\chicon(\mu) = c_0 + c_1 \tanh [c_2(\mu - c_3)]$ with $c_0, c_1$ quadractic in $T$ and $c_2, c_3$ quadractic in $1/T$. These interpolation are plotted  along the numerical results.
  
 
Our results for the $L^2=6^2$ lattices are shown in Fig.~\ref{plt:6}. The distinctive feature is the rapid transition from $\chicon\gg0$ to $\chicon\approx0$ as $\mu$ increases. As expected on physical grounds, the transition sharpens with lowering $T$. 
We present the phase diagram of $\chicon$ in the $(T,\mu)$ plane in the right panel of Fig.~\ref{plt:7}.  The heat map is the smooth interpolation of our results based on the fit discussed above.  As expected, $\chicon\approx0$ at large values of $T$ or $\mu$.  To estimate the location of the transition from a chirally broken to a chirally restored phase we have highlighted the contour at $\chicon_{\mu,T}=0.5\chicon_{0}$. 
 
A natural question is whether the transition between these two regimes is a true phase transition. 
Since chiral symmetry is explicitly broken by $m_f$, we do not expect a second order transition line, but a first order transition could exist at  small $T$ and large $\mu$. 
An indication of a true phase transition would be the sharpening of the transition as the volume grows.
In the left panel of Fig.~\ref{plt:7} we show $\chicon$ as a function of~$\mu$ for $\beta = 10$ and $L^2=6^2,10^2$.
The $\chicon$ transition indeed sharpens with $L^2$ but the data we presently have does not allow a definitive answer on whether this extrapolates to a genuine transition at infinite volume. 

In this work, we have extended the sign-optimized manifold method to reduce the finite-density sign problem of a $(2+1)$d field theory. The integration manifold was chosen by maximizing $\asi$ over a family of manifolds for which fast hybrid Monte Carlo calculations are possible. The speed at which independent configurations can be collected compensates for the still substantial sign problem on the family of manifolds. Using this method, calculations on lattice sizes up to $10^3$ and $12\times6^2$ were feasible. These calculations were enough to outline the broad features of the system's phase diagram. We find a low temperature/density region with a large chiral condensate and a high temperature/density region where the condensate is very small. Investigation of the detailed nature of the phase transition is saved for future work.

It is likely that other manifolds providing a better compromise between speed of calculation and average sign exist and can be found. Greater analytical insight into the geometry of complexified field theories could yield such manifolds. 
Another direction for future research is the extension of our methods to gauge theories.  Although the general idea of changing the domain of integration is shown to be sound~\cite{Alexandru:2018ngw}, suitable manifolds were found only through the computationally expensive method of solving the holomorphic flow equations.

\begin{acknowledgments}
A.A. is supported in part by the National Science Foundation CAREER grant PHY-1151648 and by U.S. Department of Energy grant DE-FG02-95ER40907.  P.F.B., H.L., S.L., and N.C.W.  are supported by U.S. Department of Energy under Contract No. DE-FG02-93ER-40762.
\end{acknowledgments}
\bibliographystyle{apsrev4-1}
\bibliography{thimbology}
\end{document}